\documentclass[aps,prl,groupaddress,twocolumn,floatfix]{revtex4-1}
\usepackage{units}
\usepackage{amsmath}
\usepackage{amssymb}
\usepackage{graphicx}
\usepackage{bm}
\usepackage{multirow,color,relsize,ulem,microtype}

\newcommand{\be}{\begin{equation}}
\newcommand{\ee}{\end{equation}}

\newcommand{\dt}[1]{\frac{\partial #1}{\partial t}}

\newcommand{\im}[1]{\text{Im}[#1]}

\newcommand{\TA}{\tilde{A}}

\begin{document}
\title{Pseudo-Hermitian Transition in Degenerate Nonlinear Four-Wave Mixing}

\author{Li Ge}
\email{li.ge@csi.cuny.edu}
\affiliation{\textls[-18]{Department of Engineering Science and Physics, College of Staten Island, CUNY, Staten Island, NY 10314, USA}}
\affiliation{The Graduate Center, CUNY, New York, NY 10016, USA}

\author{Wenjie Wan}
\affiliation{The State Key Laboratory of Advanced Optical Communication Systems and Networks, Department of Physics and Astronomy, Shanghai Jiao Tong University, Shanghai 200240, China}
\affiliation{University of Michigan-Shanghai Jiao Tong University Joint Institute, Shanghai Jiao Tong University, Shanghai 200240, China}
\affiliation{Key Laboratory for Laser Plasmas (Ministry of Education), Shanghai Jiao Tong University, Shanghai 200240, China}

\date{\today}

\begin{abstract}
We show that degenerate four-wave mixing (FWM) in nonlinear optics can be described by an effective Hamiltonian that is pseudo-Hermitian, which enables a transition between a pseudo-Hermitian phase with real eigenvalues and a broken pseudo-Hermitian phase with complex conjugate eigenvalues. While bearing certain similarity to that in Parity-Time symmetric systems, this transition is in stark contrast because of the absence of gain and loss in the effective Hamiltonian. The latter is real after factoring out the system decay, and the onset of non-Hermiticity in degenerate FWM is due to the total phase change of the signal wave and the idler wave. This property underlines the intrinsic coherence in FWM, which opens the door to probe quantum implications of exceptional points.
\end{abstract}

\maketitle
Exceptional points (EPs) are intriguing topological structures responsible for many counter-intuitive phenomena in non-Hermitian systems \cite{EP1,EP2,EPMVB,EP3,EP4,EP5,EP6,EP8}, including the phase transition in Parity-Time ($\cal PT$) symmetric systems \cite{Bender1,Bender2,Bender3,El-Ganainy_OL06,Moiseyev,Musslimani_prl08,Makris_prl08,Kottos,Longhi,RC,Microwave,Regensburger,CPALaser,PTConservation,Robin,degPT} and laser self termination in a photonic molecule laser \cite{EP7,EP_exp,EP_CMT,Nonl_PT}. EPs occur when two or more eigenvalues and eigenstates of an effective Hamiltonian coalesce, and so far the study of EPs has been limited to complex-valued effective Hamiltonians, which can be realized, for example, by adding gain and loss to a Hermitian system. In addition, while many systems used to probe the properties of EPs are quantum in nature, the qualitative change around the EP observed so far can be explained by purely classical arguments, and quantum coherence has not been shown to play an important role.

In this letter we advance profoundly our understanding of EPs and non-Hermitian physics in general on both of these fronts, by identifying and probing a real-valued effective Hamiltonian in an intrinsically coherent process, i.e., degenerate four wave mixing (FWM). FWM is a nonlinear optical process in which two pump frequencies interact and generate two sidebands, one with higher frequency (``signal") and one with lower frequency (``idler"). The process is called degenerate FWM when the two pump frequencies coincide. The discovery of nonlinear optical effects \cite{Franken,Armstrong,Boyd} was a hallmark in the study of light-matter interactions, and FWM in particular is important for generating broadband light with high spatial coherence (i.e., supercontinuum generation) and parametric amplification.

As we show below, the effective Hamiltonian of degenerate FWM is not only real but also pseudo-Hermitian, which enables a transition between a pseudo-Hermitian phase with real eigenvalues and a broken pseudo-Hermitian phase with complex conjugate eigenvalues. While bearing certain similarity to that in $\cal PT$-symmetric systems, this transition is in stark contrast because of the absence of gain and loss in the effective Hamiltonian. As a result, this pseudo-Hermitian transition is not caused by the intensity asymmetry of the signal and idler waves in the eigenstates, but rather by the change of the total phase of these two waves. In other words, the intrinsic coherence in degenerate FWM is the key for the onset of non-Hermiticity and brings the system to the pseudo-Hermitian phase, which opens the door to probe quantum implications of exceptional points.

\begin{figure}[b]
\includegraphics[clip,width=\linewidth]{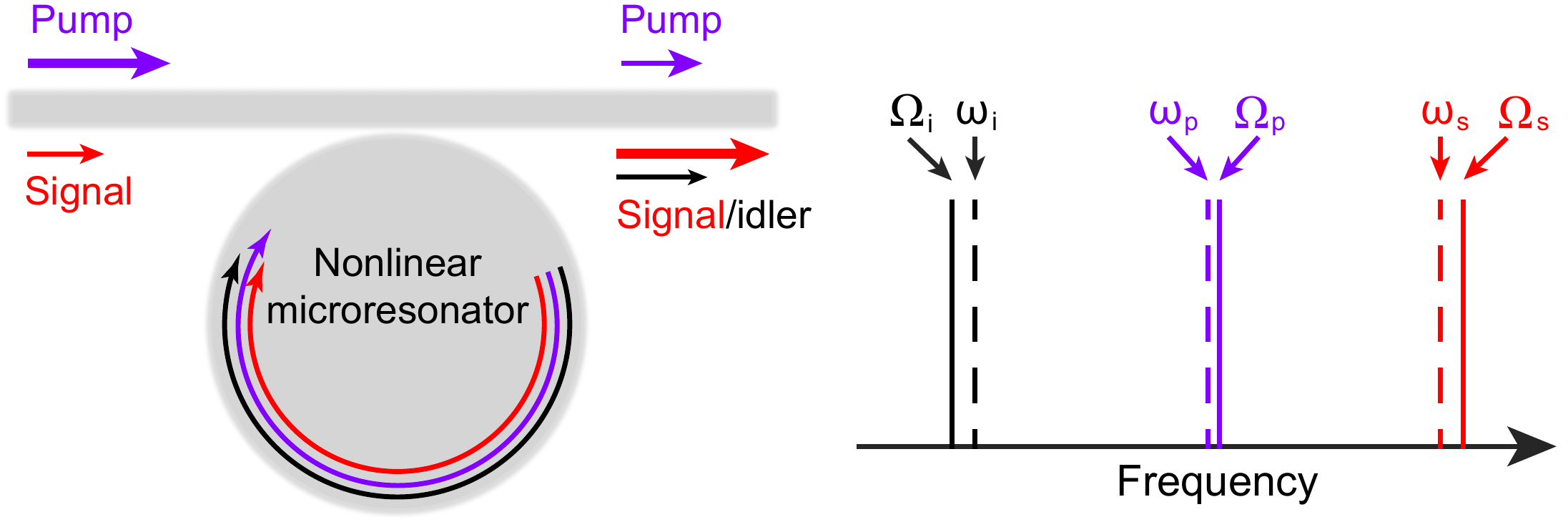}
\caption{Schematics of degenerate FWM in a microresonator. Left: Pump wave in the waveguide amplifies the input signal and generates the idler wave, via the degenerate FWM process in the nonlinear microresonator. Right: Dashed lines mark the passive resonances $\omega_i$, $\omega_p$ and $\omega_s$ in the microresonator. Solid lines mark the idler, pump, and signal frequencies from left to right.}\label{fig:schematics}
\end{figure}

Below we consider a system with passive resonances $\omega_i$, $\omega_p$, and $\omega_s$ before the pump beam is applied, with $\omega_i<\omega_p<\omega_s$ (see Fig.~\ref{fig:schematics}). These resonances can be those of an atomic system \cite{Xiao} or an engineered microstructure \cite{OIT}. A pump beam of frequency $\Omega_p$ is injected into the nonlinear medium near $\omega_p$, and a weak signal wave of frequency $\Omega_s$ and amplitude $A_s^{in}$ may also be applied near $\omega_s$ to probe the response of the system. We will refer to the other generated sideband at frequency $\Omega_i\sim\omega_i$ as the idler wave. 

Such a degenerate FWM process can be described by the following coupled equations \cite{Xiao}:
\begin{align}
i\dt{A_s} &= (-\Delta_s - i\kappa_{s})A_s - pA_s^{in} - gA_i^*,\label{eq:FWM1}\\
i\dt{A_i} &= (-\Delta_i - i\kappa_{i})A_i - gA_s^*. \label{eq:FWM2}
\end{align}
%We have written these equations in the form of the Schr\"odinger equation, with the imaginary unit $i$ in front of the time derivative on the left hand side. 
Here $A_{s,i}$ are the slowly-varying (complex) amplitudes of the signal wave and the idler wave, measured in their respective rotating frames and inside the nonlinear medium. Their decay rates are denoted by $\kappa_{s,i}$ and for simplicity we consider $\kappa_s=\kappa_i\equiv\kappa$. $\Delta_{s,i}$ are the frequency detunings from the corresponding passive frequencies, i.e. $\Delta_{s,i}=\omega_{s,i}-\Omega_{s,i}+G_{s,i}$ \cite{OIT}. We note that $G_{s,i}$ are the nonlinear frequency shifts due to the coupling of the pump beam and the signal/idler wave (i.e., ``cross phase modulation"), and they are proportional to the intensity of the pump beam. This linear relation holds for the coupling constant $g$ as well, which we take to be real and positive for convenience. As we will see, this choice does not change the property of our system. Finally, $p$ is the coupling rate of the injected signal wave into the system.

To determine the inherent dynamics of degenerate FWM, we first study the effective Hamiltonian $H$ of the system in the absence of the signal input (i.e., $A_s^{in}=0$). By performing a negative complex conjugation of Eq.~(\ref{eq:FWM2}), we arrive at the following equations
for $(\TA_s,\TA_i^*)^T\equiv[A_s\exp(\kappa t), A_i^*\exp(\kappa t)]^T$:
\begin{align}
\hspace{-2mm}
i\dt{}
\begin{pmatrix}
\TA_s\\
\TA_i^*
\end{pmatrix}
\hspace{-2pt}
=
\hspace{-2pt}
H
\begin{pmatrix}
\TA_s\\
\TA_i^*
\end{pmatrix},
\hspace{-2pt}
\;
H
\hspace{-2pt}
\equiv
\hspace{-2pt}
\begin{pmatrix}
-\Delta_s & -g\\
g & \Delta_i
\end{pmatrix}\label{eq:H}
\hspace{-2pt}
.
\end{align}
Note that $H$ is real without explicit gain and loss terms, and ``T" denotes the matrix transpose.
Now with a finite signal input $A_s^{in}=0$, the time dependence of the system is given by
\begin{align}
\begin{pmatrix}
A_s(t)\\
A_i^*(t)
\end{pmatrix}
 &= \begin{pmatrix}
A_{s0}\\
A_{i0}^*
\end{pmatrix} + \sum_{m=\pm}\alpha_m(t) |\phi_m\rangle, \label{eq:dyna}\\
\alpha_m(t) &= \alpha_m(0) e^{-i\lambda_m t-\kappa t},\nonumber
\end{align}
where $\lambda_\pm$ and $|\phi_\pm\rangle$ are the eigenvalues and eigenstates of $H$, 
$\alpha_\pm(0)$ are the projections of the initial state onto $|\phi_\pm\rangle$, and $A_{s0},A_{i0}$ are the steady state solutions of Eqs.~(\ref{eq:FWM1}) and (\ref{eq:FWM2}), obtained by setting their time derivatives to zero:
\begin{align}
A_{s0} &= -\frac{p(\Delta_i-i\kappa)}{(\Delta_s+i\kappa)(\Delta_i-i\kappa)-g^2}A_s^{in}, \label{eq:SS1}\\
A_{i0} &= -\frac{g}{\Delta_i+i\kappa}A_{s0}^*.\label{eq:SS2}
\end{align}

We note that although $H$ given by Eq.~(\ref{eq:H}) is real, it does not always have real eigenvalues as mentioned earlier. In fact it is pseudo-Hermtian \cite{PseudoH}, which is defined by
\be
H^\dagger = \sigma H \sigma^{-1} \neq H \label{eq:PseudoH}
\ee
in general, where $\sigma$ is an arbitrary matrix, $\sigma^{-1}$ is its inverse, and ``$\dagger$" denotes Hermitian conjugation. For a pseudo-Hermitian matrix $H$, its eigenvalues $\lambda$ can either be real or form complex conjugate pairs \cite{PseudoH}. The first scenario requires $\sigma^{-1}|\phi_m\rangle\propto|\phi_m\rangle$, where $|\phi_m\rangle$ is an eigenstate of $H$; the second scenario implies that $\sigma^{-1}|\phi_m\rangle\propto|\phi_n\rangle$, where $|\phi_n\rangle$ is a different eigenstate and the corresponding eigenvalues satisfy $\lambda_m=\lambda_n^*$.

\begin{figure}[b]
\includegraphics[clip,width=\linewidth]{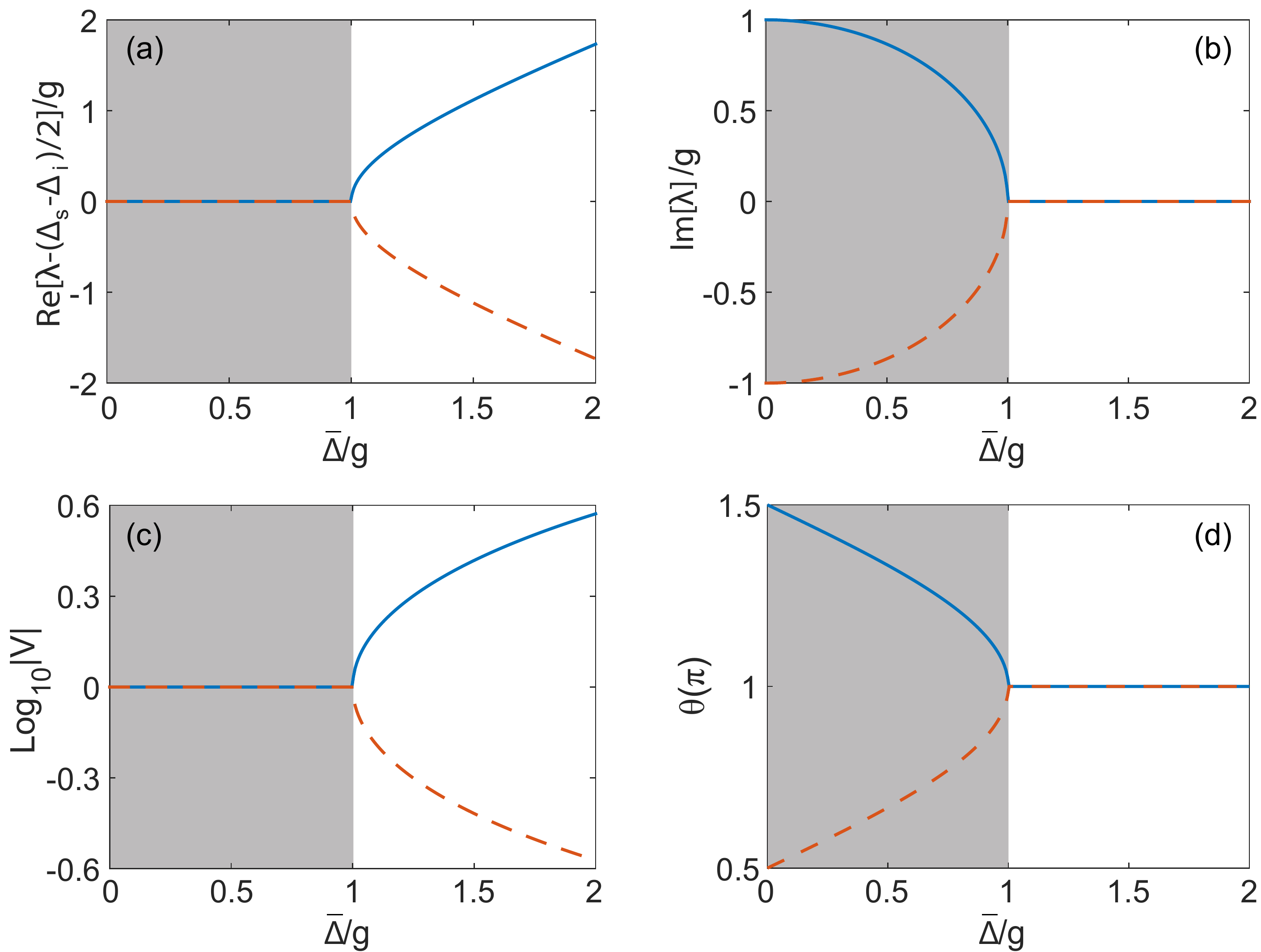}
\caption{(Color Online) Pseudo-Hermitian transition in degenerate FWM. (a,b) Real and imaginary part of the eigenvalues given by Eq.~(\ref{eq:lambda}). (c,d) Relative amplitude and total phase of the signal and idler wave in the corresponding eigenstates of $H$. $V$ is defined by $\TA_s/\TA_i^*$ and $\theta=\text{Arg}[V]=\text{Arg}[A_sA_i]$. The broken pseudo-Hermitian phase is marked by the shadowed area. $g$ is taken to be positive here.}\label{fig:eig}
\end{figure}

For degenerate FWM described by Eq.~(\ref{eq:H}), the transformation $\sigma$ in Eq.~(\ref{eq:PseudoH}) is given by the Pauli matrix $\sigma_z=\left(\begin{smallmatrix} 1 & 0 \\ 0 & -1 \end{smallmatrix}\right)$, with $\sigma_z^{-1}=\sigma_z$. The two eigenvalues of $H$ are given by
\be
\lambda_\pm = \frac{\Delta_i-\Delta_s}{2} \pm \sqrt{\bar\Delta^2-g^2},\label{eq:lambda}
\ee
where the EP is located at $|\bar\Delta|\equiv|\Delta_s+\Delta_i|/2=g$. $\lambda_\pm$ are both real when $|\bar\Delta|\geq g$, which we will refer to as the pseudo-Hermitian phase; they form a complex conjugate pair when $|\bar\Delta|<g$, which we will refer to as the broken pseudo-Hermitian phase [see Fig.~\ref{fig:eig}(a,b)]. We note that the two (off-diagonal) coupling elements of $H$ have opposite signs. This is a distinct feature of degenerate FWM and leads to the pseudo-Hermitian transition; $H$ would not have an EP or complex eigenvalues if its real-valued coupling elements have the same sign, e.g., $H_{12}=H_{21}$ (or more generally, $H_{12}=H_{21}^*$ if they are complex). The choice of a positive $g$ we have taken does not affect the properties of $\lambda_\pm$, because $H_{21}$ becomes $g^*$ if $g$ is complex, and the $g^2$ term in the radicand of Eq.~(\ref{eq:lambda}) is simply replaced by $|g|^2$.

To distinguish this pseudo-Hermitian transition from a similar one in $\cal PT$ symmetric systems and anti-$\cal PT$ symmetric systems \cite{antiPT}, first it is worth noting in all these systems the transition occurs at a square root singularity, where the radicand vanishes and marks an EP. In $\cal PT$ symmetric systems, the transition can be understood straightforwardly from the change of their spatial intensity profiles: when the optical modes have equal intensities in the gain and loss regions, they do not feel any net gain or loss, and hence the eigenvalues are real; when this symmetry is broken, then the mode with a higher intensity in the gain (loss) region will be subjected to a net gain (loss), resulting in an amplifying (decaying) eigenvalue with a positive (negative) imaginary part. This is a purely classical effect and quantum coherence does not play a role.

In degenerate FWM, there is no gain or loss in the effective Hamiltonian given by Eq.~(\ref{eq:H}), and attempting to interpret the pseudo-Hermitian transition in a similar way, i.e., using the intensity asymmetry of the signal wave and the idler wave, is problematic. In fact, we find that in the pseudo-Hermitian phase the two waves have \textit{different} intensities in both eigenstates of $H$ and that in the broken pseudo-Hermitian phase they have the same intensity [see Fig.~\ref{fig:eig}(c)]. These observations are exactly \textit{opposite} of $\cal PT$-symmetric systems, which implies that the pseudo-Hermitian transition must be driven by a different mechanism.

To identify this mechanism, we note that FWM is an intrinsically coherent process, and the phases of the signal wave and the idler wave play an important role. This feature is first manifested by a sudden change of $\theta\equiv\text{Arg}[A_sA_i]=\text{Arg}[\tilde{A}_i\tilde{A}_s]$ across the EP [see Fig.~\ref{fig:eig}(d)]: for $|\bar\Delta|\geq g$, the phase $\theta$ in both eigenstates of $H$ is independent of $|\bar\Delta|/g$ and equal to $\pi$ (0) when $g$ is positive (negative); for $|\bar\Delta|<g$ however, $\theta$ evolves differently in these two eigenstates as a function of $|\bar\Delta|/g$. To further confirm that the change of $\theta$ is the factor driving the pesudo-Hermitian transition, we write down the dynamical equation for $|\tilde{A}_s|^2+|\tilde{A}_i|^2$ when there is no signal input (i.e., $A_s^{in}=0$):
\be
\frac{\partial}{\partial t}\left(|\tilde{A}_s|^2+|\tilde{A}_i|^2\right) = 2g\im{\tilde{A}_i\tilde{A}_s}.\label{eq:energy}
\ee
The right hand side of this equation, which equals $2g|\tilde{A}_i\tilde{A}_s|\sin\theta$, highlights the role of the coherence represented by $\theta$. When $\theta=0$ or $\pi$, the system does not exhibit additional loss or gain and stays in the pseudo-Hermitian phase. When $\theta$ changes its value, energy flows in or out from the system via the pump and the system reaches the broken pseudo-Hermitian phase. These behaviors are exactly what Fig.~\ref{fig:eig} depicts.
%We note that the qualitative change of the relative phase in a pseudo-Hermitian transition was first reported in the $\cal PT$-symmetry breaking of the scattering matrix \cite{PTConservation}, which however was a byproduct rather than the origin of the pseudo-Hermitian transition in that case; the change of the mode profiles explained above lies at the heart of the transition there.

As we have shown, the average detuning $\bar\Delta$ and the nonlinear coupling $g$ are the controling knobs to tune the systems across the exceptional point. $g$ can be conveniently varied by changing the pump intensity, and so does $\bar\Delta$ via the nonlinear frequency shifts $G_{s,i}$. The phase matching condition of degenerate FWM requires $2\bm{k}_p = \bm{k}_i + \bm{k}_s$, where $\bm{k}_{i,p,s}$ are the wavevectors of the idler, pump, and signal waves.
Assuming a simple linear dispersion relation $\Omega\propto |k|$ as well as $\bm{k}_i \parallel \bm{k}_p \parallel \bm{k}_s$ and equally spaced passive resonances (i.e., $2\omega_p=2\omega_i+\omega_s$), we find that $\bar\Delta = (G_s + G_i)/2\approx 2g$ when the pump frequency coincides with the central resonance ($\Omega_p=\omega_p$) \cite{OIT}, which indicates that the system is always in the pseudo-Hermitian phase with $\im{\lambda_\pm}=0$. To reduce the value of $\bar\Delta$ to $g$ and reach the broken pseudo-Hermitian phase, one possibility is to blue-shift the pump beam from the center resonance, i.e., $\Omega_p>\omega_p$. As a result, $\Omega_i+\Omega_s=2\Omega_p$ still holds but the sum is larger than $2\omega_p=\omega_s+\omega_i$, meaning that $|\bar\Delta|\approx |2g-(\Omega_p-\omega_p)|<2g$. This observation indicates that for a given pump power (and hence a given $g$ as well), the system enters the broken pseudo-Hermitian phase when the pump frequency $\Omega_p$ becomes larger than $\omega_p+g$. Alternatively, by fixing the frequency of the blue-detuned pump beam and reducing its intensity (and hence $g$), the system enters the broken pseudo-Hermitian phase when $g$ becomes smaller than $\Omega_p-\omega_p$.

To experimentally detect the pseudo-Hermitian transition, we first note that both eigenstates of the system, expressed in terms of $(A_s,A_i^*)^T$ and with the system decay $\kappa$ restored, exhibit a decaying behavior near the EP [see Eq.~(\ref{eq:dyna})]. Therefore, with a finite signal input $A_s^{in}\neq 0$, the signal wave and idler wave approach their steady state values given by Eqs.~(\ref{eq:SS1}) and (\ref{eq:SS2}) in the long run, which however do not display a qualitatively change across the EP. To overcome this inconvenient, we propose to observe the pseudo-Hermitian transition in the transient regime and without the input signal.

\begin{figure}[t]
\includegraphics[clip,width=\linewidth]{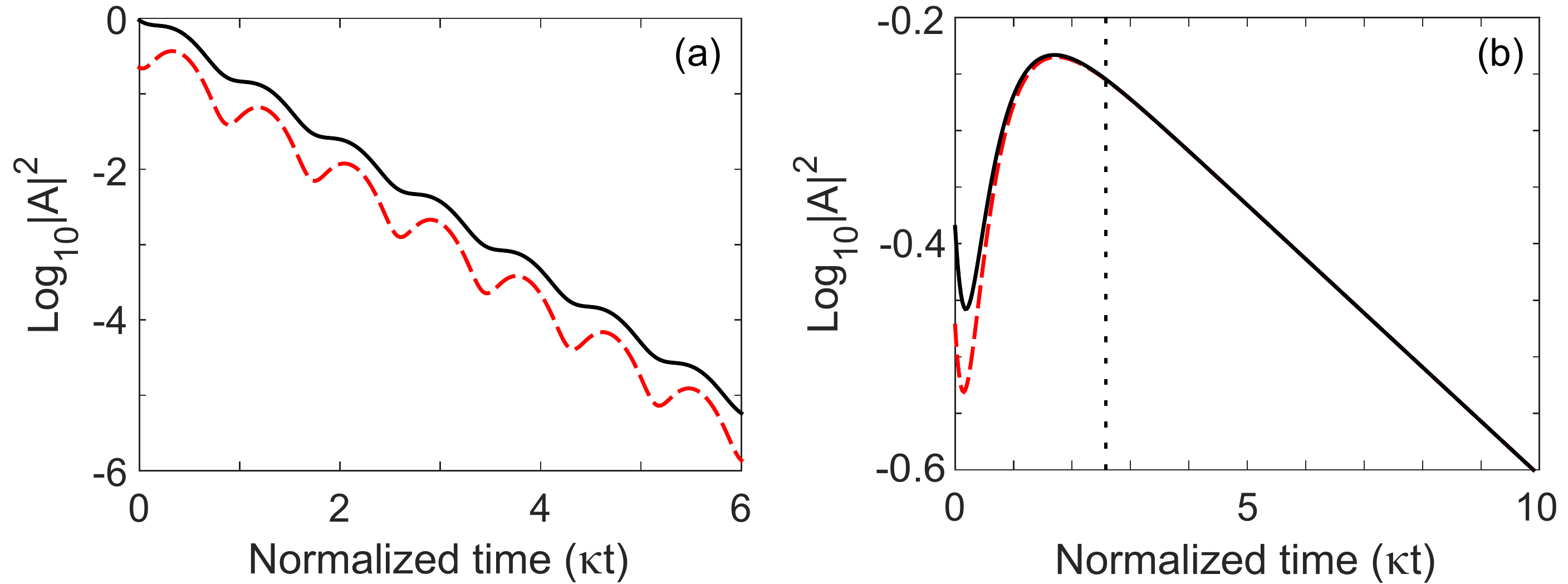}
\caption{(Color Online) Transient behaviors in the pseudo-Hermitian phase (a) and the broken phase (b) in degenerate FWM without the probe signal $A_{s}^{in}$. The dashed and solid lines show the signal and idler waves, respectively. In (a) $\Delta_s/g=5$, $\Delta_i/g=0.5$, and in (b) $\Delta_s/g=3$, $\Delta_i/g=-1.5$. The vertical dotted line in (b) is at $\kappa t=5\kappa/(\kappa+\sqrt{g^2-{\bar\Delta}^2})\approx2.57$. Arbitrary initial values of $A_s$ and $A_i$ are chosen in both (a) and (b) and $\kappa/g=0.7$. }\label{fig:dyna}
\end{figure}

The transient behavior can be probed in a ring-down measurement at either the signal frequency or the idler frequency. If we measure the intensity of these waves, then the common term of the eigenvalues $\lambda_\pm$, i.e., $(\Delta_s-\Delta_i)/2$, is eliminated. Therefore, $|A_{s,i}(t)|^2$ display a beating frequency $|\lambda_+-\lambda_-|=2\sqrt{{\bar\Delta}^2-g^2}$ while decaying at the rate $2\kappa$ in the pseudo-Hermitian phase, i.e., they are underdamped [see Fig.~\ref{fig:dyna}(a)]. A unlikely exception is when $(A_s,A_i^*)^T$ happens to be one eigenstate of $H$, then the beating disappears and $|A_{s,i}(t)|^2$ decay monotonically at the rate $2\kappa$.

In the broken pseudo-Hermitian phase close to the EP, if $(A_s,A_i^*)^T$ happens to be one eigenstate of $H$, then $|A_{s,i}(t)|^2$ decay monotonically at the same rate given by $2(\kappa\pm\sqrt{g^2-{\bar\Delta}^2})$, where the sign is determined by the eigenstate. Again this scenario is unlikely, and in general the value of $(A_s,A_i^*)^T$ at the beginning of the ring-down measurement is a superposition of the two eigenstates of $H$. Then depending on the superposition, one of $|A_{s,i}(t)|^2$ or both of them can display an anomalous transient amplification [see Fig.~\ref{fig:dyna}(b)]. This counter-intuitive feature is a manifestation of the non-Hermicity of $H$ and the nonorthogonality of the two eigenstates \cite{Makris_PRX14}. After the component $|\alpha_-(t)|^2$ in Eq.~(\ref{eq:dyna}) becomes much smaller than $|\alpha_+(t)|^2$ due to its faster decay rate $2(\kappa+\sqrt{g^2-{\bar\Delta}^2})$, $|A_{s,i}(t)|^2$ decay at the slower rate $2(\kappa-\sqrt{g^2-{\bar\Delta}^2})$ monotonically. The absence of beating then indicates that the system has entered the broken pseudo-Hermitian phase.
If the system is deep in the pseudo-Hermitian phase, the aforementioned rate $2(\kappa-\sqrt{g^2-{\bar\Delta}^2})$ becomes negative, meaning that the signal wave and the idler wave are generated and amplified without an input signal. This is exactly where supercontinuum generation mentioned in the introduction takes place, and the growth of the generated waves stops when the pump becomes saturated.

In conclusion, we have identified and probed a pseudo-Hermitian transition in degenerate FWM, which can be described by a real-valued effective Hamiltonian. The intrinsic coherence in this nonlinear optical process distinguishes such a transition from that in $\cal PT$-symmetric systems: the latter is caused by an asymmetric spatial profile of the eigenstates, which concentrates in either the gain part or the loss part of the system to exhibit an amplifying or decaying behavior. This behavior is classical in nature, and the quantum coherence does not play a role. In contrast, it is the total phase of the signal wave and the idler wave that drives the pseudo-Hermitian transition in degenerate FWM, which highlights the importance of coherence in this process and opens the door to probe quantum implications of exceptional points. This new degree of freedom holds the promise of extending our understanding of non-Hermitian physics in general, and in a related work it was shown that a coherent coupling in a $\cal PT$-symmetric system can lead to an anomalous $\cal PT$-transition away from an exceptional point \cite{nonEP}. Experimental realizations in these directions are currently being conducted under the supervision of the current authors, and preliminary findings have shown good agreement with the theoretical predictions. A complete analysis with theory and experimental data will be reported in the near future.

L.G. acknowledges partial support by NSF Grant No. DMR-1506987. W.W. acknowledges support by the National Natural Science Foundation of China (Grants No. 11304201, No. 61125503 and No. 61235009), the National 1000-Plan Program (Youth), and the Shanghai Pujiang Talent Program (Grant No. 12PJ1404700).

\end{document}